%% file: main.tex
\documentclass[
twocolumn,
]{ceurart}

\usepackage{listings}
\input{macros.tex}
\usepackage{pgfplots}

\begin{document}

\copyrightyear{2024}
\copyrightclause{Copyright for this paper by its authors.
  Use permitted under Creative Commons License Attribution 4.0
  International (CC BY 4.0).}

\conference{Joint Workshops at 50th International Conference on Very Large Data Bases (VLDBW'24) ---
  TaDA'24: 2nd International Workshop on Tabular Data Analysis, August 26--30, 2024, Guangzhou, China}

\title{Schema Matching with Large Language Models: an Experimental Study}


\author[1,2]{Marcel Parciak}[%
  orcid=0000-0002-6950-929X,
  email=marcel.parciak@uhasselt.be,
]
\cormark[1]

\author[2]{Brecht Vandevoort}[%
  orcid=0000-0001-7212-4625,
  email=brecht.vandevoort@uhasselt.be,
]
\author[2]{Frank Neven}[%
  orcid=0000-0002-7143-1903,
  email=frank.neven@uhasselt.be,
]
\author[1,2]{Liesbet M. Peeters}[%
  orcid=0000-0002-6066-3899,
  email=liesbet.peeters@uhasselt.be,
]
\author[1]{Stijn Vansummeren}[%
  orcid=0000-0001-7793-9049,
  email=stijn.vansummeren@uhasselt.be,
]

\address[1]{UHasselt, BIOMED, Agoralaan, 3590 Diepenbeek, Belgium}
\address[2]{UHasselt, Data Science Institute, Agoralaan, 3590 Diepenbeek, Belgium}

\cortext[1]{Corresponding author.}

\begin{abstract}
  Large Language Models (LLMs) have shown useful applications in a variety of
  tasks, including data wrangling. In this paper, we investigate the use of an
  off-the-shelf LLM for schema matching. Our objective is to identify semantic
  correspondences between elements of two relational schemas using only names
  and descriptions. Using a newly created benchmark from the health domain, we
  propose different so-called task scopes. These are methods for prompting the
  LLM to do schema matching, which vary in the amount of context information
  contained in the prompt. Using these task scopes we compare LLM-based schema
  matching against a string similarity baseline, investigating matching quality,
  verification effort, decisiveness, and complementarity of the approaches.  We
  find that matching quality suffers from a lack of context information, but
  also from providing too much context information. In general, using newer LLM
  versions increases decisiveness. We identify task scopes that have acceptable
  verification effort and succeed in identifying a significant number of true semantic matches.
   Our study shows that LLMs have potential in bootstrapping the
  schema matching process and are able to assist data engineers in speeding up
  this task solely based on schema element names and descriptions without the
  need for data instances.
\end{abstract}

\maketitle

\section{Introduction}
\label{sec:introduction}
\input{sections/introduction.tex}

\section{Methods}
\label{sec:methods}
\input{sections/methods.tex}

\section{Results}
\label{sec:results}
\input{sections/results.tex}

\section{Conclusion}
\label{sec:conclusion}
\input{sections/conclusion.tex}

\begin{acknowledgments}
  S. Vansummeren was supported by the Bijzonder Onderzoeksfonds (BOF) of Hasselt
  University under Grant No. BOF20ZAP02. This research received funding from the
  Flemish Government under the “Onderzoeksprogramma Artificiële Intelligentie
  (AI) Vlaanderen” programme. This work was supported by Research
  Foundation—Flanders (FWO) for ELIXIR Belgium (I002819N). The resources and
  services used in this work were provided by the VSC (Flemish Supercomputer
  Center), funded by the Research Foundation - Flanders (FWO) and the Flemish
  Government.
\end{acknowledgments}

\bibliography{biblio}

\end{document}

%% file: macros.tex
\usepackage{amsmath}
\usepackage{subcaption}
\usepackage{xspace}
\usepackage{xcolor}
\usepackage{colortbl}

\newcommand{\baseline}{\textit{sim}_{NG}}

\definecolor{heatmappurple}{rgb}{0.6, 0.439, 0.671}
\definecolor{heatmapgreen}{rgb}{0.106, 0.471, 0.216}

%% file: sections/introduction.tex
\emph{Schema matching}~\cite{DBLP:journals/vldb/RahmB01} constitutes a core
task in data integration \cite{doanPrinciplesDataIntegration2012}. It refers to
the problem of identifying semantic correspondences between elements of two
relational schemas that represent the same real-world concept. For example, a
schema matching system may conclude that an attribute \texttt{admittime} from
one table in a medical information system semantically corresponds to an
attribute \texttt{visit\_start\_date} in another table. Once correspondences
are identified, they can be used to translate data from the source schema into
data conforming to the target schema \cite{doanPrinciplesDataIntegration2012},
a process known as \emph{schema mapping}. In this paper, we focus on schema
matching.

Schema matching systems are software systems that help data engineers perform
schema matching. They generate a set of \emph{match candidates} (i.e., candidate
correspondences) which the data engineer can accept, reject or edit in order to
obtain a final set of matches \cite{DBLP:journals/vldb/RahmB01}. To generate
match candidates, a wide variety of signals that hint at element correspondence
have been considered in the research literature. These include syntactic
similarity of attribute names; consulting thesauri; looking at data values and
their distributions in concrete database instances; and exploiting database
constraints~\cite{DBLP:journals/vldb/RahmB01, DBLP:journals/pvldb/BernsteinMR11,
  chenBigGorillaOpenSourceEcosystem2018,
  asif-ur-rahmanSemiautomatedHybridSchema2023}. Unfortunately, many such signals
remain unavailable in real-world
schemas~\cite{mukherjeeLearningKnowledgeGraph2021}: attribute names are often
cryptic and involve domain-specific abbreviations not occurring in thesauri.
Use of actual data values and concrete database instances may be restricted for
legal reasons; e.g., this is the case in the health domain where real database
instances are problematic to obtain due to privacy constraints. In the absence of
available real instances, one may consider leveraging
synthetically-generated instances to aid in matching. However, accurately
replicating the complexity and subtle patterns of real medical data is highly
challenging and time-consuming, and rigorous validation of generated schema
matches is hindered by the lack of a true ground truth. In the health domain
setting, it is hence vital to be able to generate match candidates with as
little information as possible.

In the healthcare data integration context, we found that, despite its
restrictions, we often have schema documentation in the form of data
dictionaries available, as well as natural-language descriptions of some schema
elements. In particular, target schemas are often \emph{common data models}:
data schemas designed by community consensus that harmonize healthcare data
\cite{observationalhealthdatasciencesandinformaticsBookOHDSIObservational2019}.
These data models are well documented, explaining the semantics of schema
elements in detail.

In this paper, we aim to exploit this information and present an experimental
study on schema matching using an off-the-shelf generative \emph{Large Language
  Model} (LLM). We investigate how LLMs can be prompted to generate a set of match candidates. We focus on the use of schema documentation
as the sole signal and evaluate the performance against a newly defined
real-world benchmark.

We define different \emph{task scopes} for doing LLM-based schema
matching. Task scopes are prompting methods they vary primarily in the amount of
context information contained in the prompt. Using these task scopes, we aim to
answer the following research questions:
\begin{enumerate}
  \item How does the quality of LLM-based schema matching vary among different task scopes and LLM models, and how does it compare to a string-similarity-based baseline?
  \item {How decisive are LLMs in expressing their opinions on attribute pairs, and how does this affect their reliability and consistency?}

  \item What is the extent of the complementarity between the match results for different task scopes and the baseline?

  \item Is it useful and practical to combine different LLM-based and/or string-similarity-based matchings? 


\end{enumerate}

To answer these questions, we introduce the schema matching task and
experimental setup in Section~\ref{sec:methods}. There, we also introduce the different task scopes. We then present and discuss our findings w.r.t.\ the first two research questions in Section~\ref{sec:quality_of_results} and investigate the last two questions in Section~\ref{sec:combine}.  We conclude in Section~\ref{sec:conclusion}.


\paragraph*{Related Work.}%
\label{sec:related_work}




We traditionally perform schema matching by exploiting
signals such as syntactic similarity of attribute names; thesauri; data values
and distributions; and database constraints~\cite{DBLP:journals/vldb/RahmB01,
  DBLP:journals/pvldb/BernsteinMR11, chenBigGorillaOpenSourceEcosystem2018,
  asif-ur-rahmanSemiautomatedHybridSchema2023}. In this work, we are interested
in schema matching using LLMs in more restricted settings where, except for
schema documentation (i.e., the attribute names and their natural-language
descriptions) these signals remain unavailable.

\emph{Dataset discovery} is the process of navigating a collection of data
sources in order to find datasets that are relevant for a task at hand, as
well as the relationships among those datasets.  It has been observed that
schema matching is a critical component in dataset discovery, and that many
dataset discovery systems implement their own schema
matcher~\cite{DBLP:conf/icde/KoutrasSIPBFLBK21,DBLP:journals/pvldb/KoutrasPSIFBK21}. Indeed,
conceptually, one can also see dataset discovery as generalizing schema
matching. Like traditional schema matching algorithms, however, dataset
discovery algorithms will aim to exploit a rich variety of signals to do the
discovery, including access to the actual data instances. By contrast, in this
work, we are interested in schema matching using LLMs in the setting where,
except for schema documentation, such signals are unavailable.

LLMs are general machine learning models trained on large and generic natural
text data, such as the web. They are able to solve a variety of tasks with no
or minimal fine-tuning effort \cite{bommasaniOpportunitiesRisksFoundation2022}.
In the field of data management, LLMs have shown promising results for data
wrangling tasks such as error detection and data imputation
\cite{narayanCanFoundationModels2022}. However, except for \cite{10184612},
they have not been widely applied to schema matching, yet.

Zhang et al~\cite{10184612} also use language models for instance-free schema
matching, but employ and fine-tune an encoder-only model (BERT). By contrast,
we use an off-the-shelf generative decoder-only model (GPT) without any need
for fine-tuning.

Also related is SMAT~\cite{zhangSMATAttentionbasedDeep2021} which uses an
attention-based neural network to match GLoVe
embeddings~\cite{penningtonGloVeGlobalVectors2014} of schema elements, but
requires a majority of the data to be labelled: 80\% of the data that needs to
be matched is used for training, and subsequently an additional 10\% is used for
tuning weights, leaving only 10\% to evaluate the model. For practical
applications, this presents a significant limitation, as requiring 90\% of the
input schemas to be labeled, amounts to almost completely matching the schemas by
hand. Our approach, however, does not require any labelled data, allowing an
off-the-shelf usage.

AdnEV~\cite{DBLP:journals/pvldb/ShragaGR20} proposes a methodology based on deep learning and weak supervision to adjust and combine different schema matching algorithms. In this work, we observe that it makes sense to combine different task scopes to achieve the greatest effectiveness. It is an interesting direction for future work whether approaches such as AdnEV can be used to make this combination even more effective.


%% file: sections/methods.tex
\paragraph{Schema Matching.}%
\label{sec:schema_matching}

For the purpose of this paper, a \emph{schema} refers to a relational schema,
i.e., a finite set of \emph{attributes}. A \emph{1:1 match} between two schemas
$S_1$ and $S_2$ is a pair $(a,b) \in S_1 \times S_2$ that is meant to indicate
that there is a semantic correspondence between attribute $a \in S_1$ and $b
	\in S_2$. Because in the schema mapping phase we should be able to
unambiguously map data values of attribute $a$ into data values of attribute
$b$ (and vice versa) we say that $(a,b)$ is a \emph{(semantically) valid 1:1
	match} if there exists an invertible function mapping values of $a$ into values
of $b$. We define \emph{schema matching} to be the problem of deriving a set of
valid 1:1 matches {between two given schemas}.\footnote{{In this paper,
			we assume that the source and target table are
			already provided, the table matching step, i.e.,
			identifying corresponding tables, has thus already been completed.}} We note that in
the literature, also matches of kind $1:m$, $n:1$ and $n:m$ exist. For example,
in a match of kind $1:m$ we may relate a single attribute $a$ in $S_1$ to a
\emph{set} of attributes $B \subseteq S_2$, meaning that the information of
$a$-values in $S_1$ will be ``distributed'' among all the attributes in $B$ and
that we need all attributes in $B$ to recover the $a$-value. A typical example
is relating \textsf{Name} in $S_1$ to $B = \{ \textsf{First name}, \textsf{Last
		name}\}$. In this paper, we restrict ourselves to 1:1 matches for two reasons.
First, this shrinks the search space significantly for possible matches, making
our experimental approach feasible even for larger schemas. Second, it allows
us to compare our results to a baseline using string similarity measures, which
are difficult to extend to $1:m$, $n:1$ or $n:m$ matches.

\paragraph{Benchmark.}%
\label{sec:benchmark}
In order to gauge the suitability of LLMs for schema matching we have created a
new benchmark, situated in the healthcare domain. We draw source schemas from the
MIMIC-IV dataset~\cite{johnsonMIMICIVFreelyAccessible2023} and target schemas
from OHDSI OMOP Common Data Model~\cite{%
	observationalhealthdatasciencesandinformaticsBookOHDSIObservational2019}. Both
are public, well-known data models in the medical informatics community. The
OHDSI community maintains an ETL process to transform data from MIMIC-IV to
OMOP~\cite{kallfelzMIMICIVDemoData}. We use this ETL specification to manually identify
all semantically valid 1:1 matches that will serve as the ground truth. That
is, we manually inspect all applied ETL transformations and derive each
attribute combination $(a,b)$ where a single value from the source attribute $a$ is
sufficient to determine the value from the target attribute $b$ and vice versa. For
example, the attribute \texttt{gender} from MIMICs \texttt{Patients} table is
mapped to both \texttt{gender\_concept\_id} and \texttt{gender\_source\_value}
of OMOPs \texttt{Person} table. Both mappings are valid 1:1 matches, as the
value in \texttt{gender} can be mapped to a valid value fit for either
attribute and vice versa. In contrast, the attribute \texttt{admittime} of
MIMICs \texttt{Admissions} table is not a valid match for
\texttt{visit\_start\_datetime} of OMOPs \texttt{Visit\_Occurrence} table, as
the ETL specification needs to combine it with another attribute to determine
the value of \texttt{visit\_start\_datetime}.

We have extracted a total of $49$ valid 1:1 matches between $7$ relations from
MIMIC-IV and $6$ relations from OMOP. In total, there are $9$ relation pairs
that contain at least one semantic match. We will refer to each of these
relation pairs as a \emph{dataset} in our benchmark. Our $9$ datasets create a
search space of $1839$ attribute pairs that contain $49$ true semantic 1:1
matches as summarized in Table~\ref{tab:datasets}. We consider all other
attribute pairs as non-matches. The schema matching problem is hence highly
imbalanced. For each (source or target) table and each attribute we extract the
name as well as a natural language description from respective documentations.
Our benchmark is publicly available in our artefacts
repository~\cite{parciakArtifactRepositorySchema}.

We acknowledge that a benchmark consisting of public datasets is probably
contained in the training data of an LLM trained on the web. As an example,
when asking ChatGPT to give a description of the attribute \texttt{dischtime}
from the \texttt{admissions} table in MIMIC-IV, the answer returned from the
model fits the description given in the official documentation of MIMIC-IV
well. This represents a limitation of our experimental setup. We argue that
although the datasets are known to the LLM, the true semantic matches to
transform data from MIMIC-IV to OMOP are not readily explicitly available: significant effort is required to extract them from the ETL scripts.

\begin{table*}[tbp]
	\small
	\centering
	\caption{\small Benchmark datasets: names of source and target tables, their respective attributes and attribute pair counts, and the number of true semantic matches.}
	\label{tab:datasets}
	\footnotesize
	\begin{tabular}[h]{l l c l c c c}
		\hline
		dataset       & Source         & |source| & target                & |target| & |pairs| & |matches| \\
		\hline
		\texttt{AdCO} & Admissions     & 16       & Condition\_Occurrence & 16       & 256     & 2         \\
		\texttt{AdVD} & Admissions     & 16       & Visit\_Detail         & 19       & 304     & 5         \\
		\texttt{AdVO} & Admissions     & 16       & Visit\_Occurrence     & 17       & 272     & 8         \\
		\texttt{DiCO} & Diagnoses\_ICD & 5        & Condition\_Occurrence & 16       & 80      & 2         \\
		\texttt{LaMe} & Labevents      & 10       & Measurement           & 20       & 200     & 10        \\
		\texttt{PaPe} & Patients       & 6        & Person                & 18       & 108     & 5         \\
		\texttt{PrDE} & Prescriptions  & 17       & Drug\_Exposure        & 23       & 391     & 6         \\
		\texttt{SeVD} & Services       & 5        & Visit\_Detail         & 19       & 95      & 5         \\
		\texttt{TrVD} & Transfers      & 7        & Visit\_Detail         & 19       & 133     & 6         \\
		\midrule
		\multicolumn{5}{r}{Total}                                                    & 1839    & 49        \\
		\hline
	\end{tabular}
\end{table*}

\begin{figure*}[tpb]
	\centering
	\includegraphics[width=0.9\textwidth]{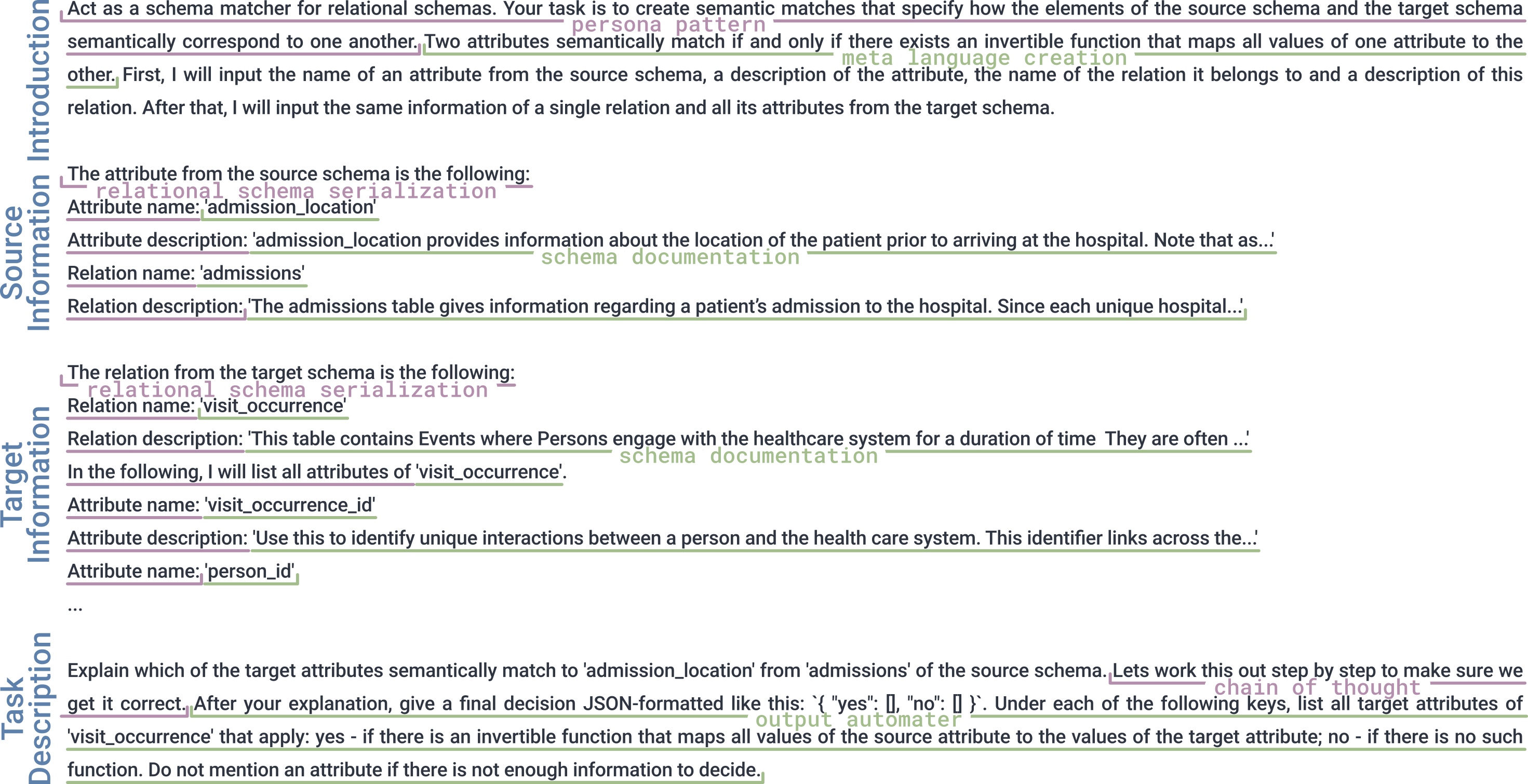}
	\caption{A truncated example of a \texttt{1-to-N} prompt. The prompt
		engineering best practices applied are highlighted.}
	\label{fig:prompt_design}
\end{figure*}

\paragraph{Prompt Engineering.}%
\label{sec:prompting_engineering}
Generative LLMs are trained to answer questions in natural language. As such,
we need to interface with the LLM via prompts that describe the task to be
performed by the LLM as well as the table and attribute names and descriptions.
Previous research into prompt engineering has proposed a number of \emph{prompt
	engineering patterns} that positively influence answer quality~\cite{%
	narayanCanFoundationModels2022}. We next discuss how we have applied these
common practices in our prompt design by means of the visual representation in
Figure~\ref{fig:prompt_design}. Each prompt is always applied to a single
source schema and a single target schema (plus their descriptions), and
consists of four sections referred to as \emph{Introduction}, \emph{Source
	Information}, \emph{Target Information}, and \emph{Task Description}.

First, we introduce the schema matching problem to the LLM by utilizing the
\emph{Persona Pattern} to let the LLM act as a schema
matcher~\cite{whitePromptPatternCatalog2023}. We then introduce our definition
of a valid 1:1 match using the \emph{Meta Language Creation}
pattern~\cite{whitePromptPatternCatalog2023}. Both patterns are illustrated in
the \emph{Introduction} section in Figure~\ref{fig:prompt_design}.

Subsequently, we serialize the schema information, including table and
attribute descriptions, using a serialization inspired by
\cite{narayanCanFoundationModels2022}. Concretely, we first serialize the source
information, followed by the target information. An example of this can be
viewed in Figure~\ref{fig:prompt_design} in the \emph{Source Information} and
\emph{Target Information} sections.

We finalize our prompts with
the task description that utilizes the phrase ``Lets think step by step''
which has been shown to increase performance by instructing the LLM to build
up a step-by-step argument in the output~\cite{DBLP:conf/nips/KojimaGRMI22} and
to which we refer as the \emph{Chain of Thought Pattern} in
Figure~\ref{fig:prompt_design}. We end the task description with the
\emph{Output Automater} pattern to instruct the model to output structured data
(in particular: JSON) for further processing~\cite{%
	whitePromptPatternCatalog2023}. Here, we ask the LLM to generate a structured
output such that we can extract $(a,b,\textit{out})$ triples, where $a$ and $b$
are attributes from the source and target schema, respectively, and
$\textit{out}$ (discussed further below) is the LLM's opinion of whether
$(a,b)$ is a semantically meaningful 1:1 match. Both patterns are illustrated
in the example prompt in Figure~\ref{fig:prompt_design} in section \emph{Task
	Description}.

During our experiments, we found that using a three-step scale for \textit{out}
works best. We ask the LLM to use \texttt{yes} for a match, \texttt{no} for a
non-match, and \texttt{unknown} if there is not enough information to decide.
We have also experimented with numerical scores, which were difficult to
interpret, and five-step scales, which were prone to hallucinations. For
example, asking for a five-step scale of \emph{no correspondence}, \emph{low
	correspondence}, \emph{medium correspondence}, \emph{high correspondence} and
\emph{very high correspondence} frequently resulted in opinions such as
\emph{low to medium correspondence}, making a reliable interpretation highly
difficult. We note that LLM output is not necessarily complete: there may be
attribute pairs $(a,b)$ for which the LLM does not give its opinion; we treat
this as \texttt{unknown}.

\paragraph{Task Scopes.}%
\label{sec:task_scopes}
In this paper, we focus on a comparison of \emph{task scopes}, which we define
as the amount of schema information contained in a single prompt. We define
four different scopes:
\begin{description}
	\item[1-to-1] Each prompt contains exactly one attribute from source
	      and one from target.
	\item[1-to-N] Each prompt contains a single attribute from the source
	      schema and $N$ attributes of the target schema, where $N$ refers to
	      the total number of attributes in the target schema.
	\item[N-to-1] Each prompt contains $N$ attributes from the source schema
	      and a single attribute from the target schema, where $N$ refers to
	      the total number of attributes in the source schema.
	\item[N-to-M] Each prompt contains $N$ attributes from the source schema
	      and $M$ attributes from the target schema, where $N$ and $M$ refer to
	      the total number of attributes in the source and target schema,
	      respectively.
\end{description}
It is worth noting that the task scope choice has implications on the
complexity to parse structured votes from the LLM output. While we expect a
single vote (e.g. \texttt{yes} or \texttt{no}) in the \texttt{1-to-1} case, an
output to the \texttt{N-to-M} task scope potentially contains $N\times M$ votes,
one for each attribute pair.

We investigate both \texttt{1-to-N} and \texttt{N-to-1} as both
scopes present very different contexts to the LLM. In the former, the LLM is
presented with all available information about the target relation while
limiting the information of the source information, and vice versa in the
latter. We found that this difference impacts the quality of the matches.

\paragraph{String Similarity Baseline.}%
\label{sec:baseline}

We aim to compare the performance of the LLM-based approaches against a
baseline based on a string similarity measure, a well-established baseline
approach in the field  of schema mapping and ontology alignment~\cite{%
	sunComparativeEvaluationString2015, cheathamStringSimilarityMetrics2013}. To
do so, we have selected edit distance-based metrics investigated by \cite{%
	sunComparativeEvaluationString2015} and \cite{%
	cheathamStringSimilarityMetrics2013} and checked for their availability in the
common Python library \texttt{textdistance}\footnote{\url{%
		https://pypi.org/project/textdistance}}. We aim to find
commonly used similarity metrics that are readily available and identified four
metrics: Jaro Winkler, Levenshtein, Monge Elkan and N-gram.
We evaluated these metrics on our benchmark by calculating the similarities between the attribute names for each attribute combination in the benchmark. It is important to note that these attribute pairs are the same as those used in our results, although we do not report dataset-specific values here.
%
%
We then generate a ranking of all
attribute pairs 
and calculate the precision and recall for
each threshold per similarity measure. Figure~\ref{fig:similarity_metrics}
displays the corresponding precision-recall curve and reveals that N-gram with
$n=3$ is the best performing metric (w.r.t. the area under curve). This string
similarity metric will therefore be used in the following as a baseline.


Specifically, to obtain the baseline we calculate the N-gram string
similarity $\baseline(a,b)$ between all possible attribute pairs in a
dataset. For each attribute name $a'$ we obtain the set of its 3-grams $a$
after padding with special characters as described by Sun et
al.~\cite{sunComparativeEvaluationString2015}. For example, the name
\texttt{admittime} is transformed into the set \{\texttt{\#\#a},
\texttt{\#ad}, \texttt{adm}, \texttt{dmi}, \texttt{min}, \texttt{int},
\texttt{ntt}, \texttt{tti}, \texttt{tim}, \texttt{ime}, \texttt{me\%},
\texttt{e\%\%}\}. For two sets $a$ and $b$, we then calculate the Dice
similarity:
\[
	\baseline(a,b) := \frac{2 \times |a \cap b|}{|a| + |b|}
\]
Since each $\baseline(a,b)$ lies in the range $[0; 1]$, this
defines an order on match candidates, with highest values appearing first. One
can either set a threshold $\theta$ to decide which similarity value is
sufficient for a match or or limit the number of matches to the top $k$ ranked
ones. Using thresholding, all pairs $\baseline(a,b) \geq \theta$
will then be output as a match, and using ranking, all pairs $\{(a_0, b_0),
	\ldots, (a_n, b_n)\}[1:k]$ where $\baseline(a_i, b_i) \geq \baseline(a_j, b_j)$
for all $i < j$ will be output as a match. We choose the former and determine a
separate threshold per dataset as follows: we consider all calculated
similarity values as thresholds and pick the threshold that achieves the best
F1-score on the dataset. We then choose $\baseline$ with this threshold as the
baseline for the considered dataset. This approach favors the baseline, as
it overestimates the capabilities of the N-gram string similarity for schema
matching. In practice, a data engineer cannot know which threshold to use.

The choice of tresholding over ranking is motivated by the fact that the output
of the LLM does not imply any ordering, we ask for a simple \texttt{yes},
\texttt{no} or \texttt{unknown} decision instead. Hence, common ranking metrics
such as \emph{recall@k} or \emph{mean reciprocal rank} cannot be applied to the
LLM results. Furthermore, we note that our approach to determine
dataset-specific thresholds is equivalent to choosing a dataset-specific $k$
that maximizes the F1-score when interpreting the $\baseline$ as a ranking.

\begin{figure}
	\centering
	\pgfplotsset{
		width=\columnwidth,
		every axis plot/.append style={thick},
	}
	\definecolor{plotly1}{RGB}{ 76, 120, 168}
	\definecolor{plotly2}{RGB}{245, 133,  24}
	\definecolor{plotly3}{RGB}{228,  87,  86}
	\definecolor{plotly4}{RGB}{114, 183, 178}
	\begin{tikzpicture}
		\begin{axis}[
				axis lines = left,
				xlabel = {Recall},
				ylabel = {Precision},
				xmin=0.0, xmax=1.0,
				ymin=0.0, ymax=0.6,
				height=4.25cm,
			]
			\addplot[plotly1] table [x=recall,y=precision] {plotdata/jaro-winkler.dat};
			\addlegendentry{\footnotesize Jaro Winkler (AUC: 0.12)}
			\addplot[plotly2] table [x=recall,y=precision] {plotdata/levenshtein.dat};
			\addlegendentry{\footnotesize Levenshtein (AUC: 0.08)}
			\addplot[plotly3] table [x=recall,y=precision] {plotdata/monge-elkan.dat};
			\addlegendentry{\footnotesize Monge Elkan (AUC: 0.04)}
			\addplot[plotly4] table [x=recall,y=precision] {plotdata/n-grams.dat};
			\addlegendentry{\footnotesize N-gram (n=3, AUC: 0.14)}
		\end{axis}
	\end{tikzpicture}
	\caption{Precision-Recall curve of different string similarity metrics,
		tested on the attribute names of our benchmark.}
	\label{fig:similarity_metrics}
\end{figure}

\paragraph{Experimental Setup.}%
\label{sec:experimental_setup}

For a fixed dataset and fixed task scope, an \emph{experiment} consists of
sending the correspoding prompt three times to the LLM. 
We extract three \emph{votes} from the responses and use majority voting to minimize
the effect of hallucinations. If an attribute pair is missing or there is a
split decision, this pair is considered
\texttt{unknown}. Each experiment is repeated five times. The results
are compared against our benchmark by means of (\emph{i}) the
\emph{F1}-score, the harmonic mean between \emph{precision} and \emph{recall}
w.r.t.\ the ground-truth semantically valid matches, and (\emph{ii}) a
\emph{decisiveness}-score, indicating the fraction of non-\texttt{unknown}
votes. We use OpenAI's \texttt{gpt-3.5-turbo-0125} and
\texttt{gpt-4-0125-preview} models with default settings,
acknowledging the fact that performance could be improved with tuning
the settings. Jupyter notebooks that we used to obtain
the results can be found in our artefacts
repository~\cite{parciakArtifactRepositorySchema}.


%% file: sections/results.tex
We next present the findings of our experimental study on schema matching using LLMs. Section~\ref{sec:quality_of_results} focuses on the quality of the schema matching results generated by the different separate task scopes, whereas Section~\ref{sec:complementarity} addresses their complementarity and the benefits of combining task scopes.

\subsection{Quality of schema matching}
\label{sec:quality_of_results}
\input{sections/results_quality.tex}

\subsection{Complementarity} 
\label{sec:complementarity}
\label{sec:combine}
\input{sections/complementarity.tex}

%% file: sections/results_quality.tex
We begin by evaluating the quality of schema matching results produced by the different task scopes, using F1-scores for comparison both among the LLMs and against the baseline (Section~\ref{sec:F1-scores}). Next, we assess the decisiveness of the LLMs in their opinions on attribute pairs in Section~\ref{sec:decisiveness_of_results}. Finally, we analyze the consistency of our experiments across various task scopes and datasets in Section~\ref{sec:consistency}, reporting the standard deviation of F1-score, precision, and recall to illustrate the expected variance when using LLMs for schema matching.

\begin{table*}[tbp]
    \centering
    \caption{Median F1-scores, coloured for comparison against the N-gram
        similarity baseline. Green indicates an F1-score higher, purple an
        F1-score lower than the baseline. After each F1-score, in parenthesis $(p,r)$, we
        give the precision $p$ and the recall $r$. The best F1-score of each dataset is
        set in bold.
        \label{tab:f1_scores}}
    \footnotesize
    \setlength{\tabcolsep}{0.15em}
    \begin{tabular}{l c r c c c c r c c c}
                &             &  & \multicolumn{4}{c}{GPT-3.5} &        & \multicolumn{3}{c}{GPT-4}                                        \\
        dataset & $\baseline$ &  & 1-to-1                      & 1-to-N & N-to-1                    & N-to-M &  & 1-to-N & N-to-1 & N-to-M \\
        \midrule
        \input{tables/precision_recall_f1_scores.tex}
    \end{tabular}
\end{table*}

\subsubsection{F1-scores}
\label{sec:F1-scores}

Table~\ref{tab:f1_scores} shows the median F1-scores of each task scope
per dataset, for both LLMs that we tested. The colouring indicates
whether the F1-scores are higher (green) or lower (purple) than $\baseline$.
The best F1-score of each dataset is set in bold. We observe that the maximal
F1-scores range from $0.364$ to $0.800$, highlighting a variation in the
difficulty across the datasets. The bottom row displays the mean per column
over all datasets and reveals the following general trends:
\begin{itemize}
    \item all task scopes, except for \texttt{1-to-1}, outperform the baseline
          $\baseline$;
    \item each task scope shows an improvement in F1-score when moving from
          \texttt{GPT-3.5} to \texttt{GPT-4}; and,
    \item under the task scopes tested on both LLMs, \texttt{N-to-M} has the
          lowest mean F1-score.
\end{itemize}
Next, we conduct a more detailed analysis of each task scope in relation to the datasets.

We see that \texttt{1-to-1} is the least performing task scope: it fails to
achieve the maximal F1-score on any dataset and is worse than (or on par with)
the N-gram baseline, with the exception of the \texttt{DiCO} and the
\texttt{LaMe} datasets. Moreover, \texttt{1-to-1} is typically outperformed by
other scopes that incorporate more information in their prompts. Consequently,
we assert that it lacks sufficient information for making informed,
high-quality decisions.

Due to the low performance of \texttt{1-to-1} under \texttt{GPT-3.5}, combined
with its high monetary cost, we decided to exclude the \texttt{1-to-1} task
scope for our experiments using \texttt{GPT-4}.

For the analysis of the remainder of the task scopes, we use the following
format. For a fixed task scope, we first consider \texttt{GPT-3.5}, and compare
it with the baseline and the other task scopes run under \texttt{GPT-3.5}. We
then make a comparison with its \texttt{GPT-4} counterpart. Finally, we consider
the task scope run under \texttt{GPT-4} and compare with the baseline and the
other task scopes for \texttt{GPT-4}.

The \texttt{1-to-N} task scope obtains the maximal \texttt{GPT-3.5} F1-score on
the \texttt{DiCO}, \texttt{PrDE} and the \texttt{SeVD} dataset, outperforming
$\baseline$ on five of nine datasets. With a single exception on the
\texttt{PaPe} dataset, \texttt{1-to-N} dominates \texttt{1-to-1}. By this
comparison, we deduce that adding more context information to a single prompt
improves the quality of the LLM's decisions. The scores of \texttt{1-to-N} can
be further improved by using \texttt{GPT-4}, the \texttt{SeVD} dataset being
the single exception. This improvement can be attributed to an increase in
precision on each dataset except on \texttt{DiCO} where it remains the same.
With \texttt{GPT-4}, \texttt{1-to-N} outperforms $\baseline$ on eight datasets.

Using the \texttt{N-to-1} task scope we see improved F1-scores on average,
achieving the maximal F1-score of the baseline and all \texttt{GPT-3.5}-based
experiments four times and dominating $\baseline$ on five of nine datasets.
Further, \texttt{N-to-1} dominates \texttt{1-to-1} on all datasets except for
the \texttt{DiCO} dataset, reinforcing the deduction we made for
\texttt{1-to-N}: adding context information improves matching quality. Using
\texttt{GPT-4}, the \texttt{N-to-1} task scope dominates the N-gram baseline on
every single dataset, achieving the highest F1-score on five datasets as well
as the highest F1-score on average. Analogous to \texttt{1-to-N}, the
improvement can be attributed to an improved precision score on every dataset
while the recall decreases on one dataset.

Finally, the \texttt{N-to-M} task scope achieves a maximal F1-score among the
\texttt{GPT-3.5} approaches three times, outperforming both $\baseline$ and
\texttt{1-to-1} on five and six datasets, respectively. With the exception of
\texttt{SeVD}, the use of \texttt{GPT-4} improves the F1-scores on all datasets,
resulting in the highest F1-score on three datasets. In contrast to
\texttt{1-to-N} and \texttt{N-to-1}, recall of \texttt{N-to-M} is better or on
par with its \texttt{GPT-3.5} counterpart. With \texttt{GPT-4}, \texttt{N-to-M}
dominates $\baseline$ on six datasets. We hypothesize that the failure to
improve the number of $\baseline$-dominating datasets is due to the increase in complexity
of the output format. While it is sufficient to simply list attribute names for
\texttt{1-to-N} and \texttt{N-to-1}, we need a list of attribute pairs for
\texttt{N-to-M}. 

\paragraph{Conclusion.} We observe that for both LLMs, all task scopes, except
for \texttt{1-to-1}, outperform the baseline on average with a maximal increase of 0.245 points. However, no single task
scope consistently dominates across all datasets. Across task scopes, moving
from \texttt{GPT-3.5} to \texttt{GPT-4} increases the F1-score over all data
sets (with \texttt{SeVD} as a single exception for \texttt{1-to-N} as well as
\texttt{N-to-M}) confirming the general accepted belief that transitioning to
more advanced LLMs yields better results. Interestingly, when moving from
\texttt{GPT-3.5} to \texttt{GPT-4} the rise in F1-score is due to an increase
in precision for the task scopes \texttt{1-to-N} and \texttt{N-to-1}, while for
\texttt{N-to-M} it is due to an increase in recall (sometimes even at the
expense of a slight drop in precision). Finally, within the same LLM,
\texttt{N-to-M} is the least performing task scope of the three task scopes we
analysed on both LLMs.

\subsubsection{Decisiveness}%
\label{sec:decisiveness_of_results}
In the course of our experiments, we observed that the LLM often fails to
express an opinion on all attribute pairs requested. We summarize this behavior
in the decisiveness score shown in Table~\ref{tab:decisiveness}. This score
captures the ratio of attribute pairs that received a \texttt{yes} or
\texttt{no} vote--so not an \texttt{unknown}--to all attribute pairs per
dataset. As the name already indicates it measures how decisive the model is.
On most datasets, the following
inequality holds: $\texttt{1-to-1} > \texttt{N-to-1} > \texttt{1-to-N} > \texttt{N-to-M}.$  We clearly see that increasing the amount of information
per prompt decreases the decisiveness. With \texttt{GPT-3.5}, the
\texttt{N-to-1} task scope remains in an acceptable range, \texttt{1-to-N}
fluctuates between datasets while \texttt{N-to-M} is consistently in an
unacceptable range. The use of \texttt{GPT-4} improves the decisiveness
considerably for \texttt{N-to-1} and \texttt{1-to-N}. Interestingly, the
decisiveness of \texttt{N-to-M} does not profit from the larger model.

Given the low quality of results for \texttt{1-to-1}, the high decisiveness
indicates that using the \texttt{1-to-1} task scope makes the wrong decision
most of the time. This supports our decision to exclude \texttt{1-to-1} from
further experiments with \texttt{GPT-4}. The extremely low decisiveness of
\texttt{N-to-M}, however, may indicate that the complexity of the output could
play a major role in the low quality of its results. As previously mentioned,
the output of \texttt{N-to-M} is a list of tuples of attribute names while it
is sufficient to simply list attribute names for \texttt{1-to-N} and
\texttt{N-to-1}. 

\paragraph{Conclusion.} An increase of context information per prompt decreases
the number of attribute pairs an LLM expresses an opinion on. While this effect
can be mitigated using \texttt{GPT-4} for \texttt{1-to-N} and \texttt{N-to-1},
this is not the case for \texttt{N-to-M}.

\begin{table}[tbp]
    \centering
    \caption{Decisiveness scores (the number of attribute pairs that received a
        \texttt{yes} or \text{no} score--so not an \texttt{unknown}--relative to
        the total number of attribute pairs) per task scope and model.}
    \label{tab:decisiveness}
    \footnotesize
    \setlength{\tabcolsep}{0.25em}
    \begin{tabular}{l r c c c c r c c c}
                            &  & \multicolumn{4}{c}{GPT-3.5} &                    & \multicolumn{3}{c}{GPT-4}                                                                                        \\
        \scriptsize dataset &  & \scriptsize 1-to-1          & \scriptsize 1-to-N & \scriptsize N-to-1        & \scriptsize N-to-M &  & \scriptsize 1-to-N & \scriptsize N-to-1 & \scriptsize N-to-M \\
        \midrule
        \input{tables/decisiveness.tex}
    \end{tabular}
\end{table}

\subsubsection{Consistency}%
\label{sec:consistency}

We have been reporting results with respect to the median. Given that we
conducted each experiment five times, it is interesting to investigate the
consistency of the experiment results. We do so by reporting the standard
deviation of F1-scores, precision and recall in Table~\ref{tab:consistency}. We
see that, on average, \texttt{1-to-N} and \texttt{N-to-1} have low standard
deviations with $0.074$ and $0.062$, respectively. Both \texttt{1-to-1}
($0.141$) and \texttt{N-to-M} ($0.160$) have higher standard deviations,
\texttt{N-to-M} reaching the maximum across the whole table. Using
\texttt{GPT-4}, the results increase in consistency. \texttt{N-to-1} reaching
the overall minimum with $0.031$ followed by \texttt{1-to-N} with $0.037$.
\texttt{N-to-M} remains the least consistent but improves to $0.094$.

\paragraph{Conclusion.}
We find that the standard deviation of the F1-scores remains in acceptable
ranges ($<0.1$) for \texttt{1-to-N} and \texttt{N-to-1} on both models. With
\texttt{GPT-4}, all standard deviations improve further. We conclude that LLMs
are consistent enough to be used in practice for schema matching.

\begin{table}[tbp]
    \centering
    \caption{The standard deviation of F1-score, precision and recall (the
        latter two are presented in brackets) calculated from five experiment
        runs. A darker green indicates a lower standard deviation for the
        F1-score.}
    \label{tab:consistency}
    \footnotesize
    \setlength{\tabcolsep}{0.25em}
    \begin{tabular}{c r c r c }
		scope & & GPT-3.5 & & GPT-4 \\
        \midrule
        \input{tables/consistency_short.tex}
    \end{tabular}
\end{table}


%% file: tables/precision_recall_f1_scores.tex
AdCO & 0.286 \scriptsize (0.20, 0.50) & & \cellcolor{heatmappurple!100} 0.000 \scriptsize (0.00, 0.00) & \cellcolor{heatmappurple!53} 0.133 \scriptsize (0.08, 0.50) & \cellcolor{heatmappurple!29} 0.200 \scriptsize (0.11, 1.00) & \cellcolor{heatmappurple!0} 0.286 \scriptsize (0.20, 0.50) &  & \cellcolor{heatmapgreen!16} \textbf{0.400} \scriptsize (0.33, 0.50) & \cellcolor{heatmapgreen!16} \textbf{0.400} \scriptsize (0.25, 1.00) & \cellcolor{heatmapgreen!16} \textbf{0.400} \scriptsize (0.33, 0.50) \\
AdVD & 0.125 \scriptsize (0.07, 0.40) & & \cellcolor{heatmappurple!100} 0.000 \scriptsize (0.00, 0.00) & \cellcolor{heatmappurple!33} 0.083 \scriptsize (0.05, 0.20) & \cellcolor{heatmapgreen!14} 0.250 \scriptsize (0.16, 0.60) & \cellcolor{heatmapgreen!18} 0.286 \scriptsize (0.50, 0.20) &  & \cellcolor{heatmapgreen!14} 0.250 \scriptsize (0.18, 0.40) & \cellcolor{heatmapgreen!21} 0.316 \scriptsize (0.21, 0.60) & \cellcolor{heatmapgreen!27} \textbf{0.364} \scriptsize (0.33, 0.40) \\
AdVO & 0.333 \scriptsize (0.50, 0.25) & & \cellcolor{heatmappurple!29} 0.235 \scriptsize (0.22, 0.25) & \cellcolor{heatmappurple!3} 0.320 \scriptsize (0.24, 0.50) & \cellcolor{heatmapgreen!25} 0.500 \scriptsize (0.38, 0.75) & \cellcolor{heatmappurple!45} 0.182 \scriptsize (0.33, 0.12) &  & \cellcolor{heatmapgreen!16} 0.444 \scriptsize (0.40, 0.50) & \cellcolor{heatmapgreen!45} \textbf{0.636} \scriptsize (0.50, 0.88) & \cellcolor{heatmappurple!7} 0.308 \scriptsize (0.40, 0.25) \\
DiCO & 0.400 \scriptsize (0.33, 0.50) & & \cellcolor{heatmapgreen!44} 0.667 \scriptsize (1.00, 0.50) & \cellcolor{heatmapgreen!66} \textbf{0.800} \scriptsize (0.67, 1.00) & \cellcolor{heatmappurple!33} 0.267 \scriptsize (0.15, 1.00) & \cellcolor{heatmapgreen!44} 0.667 \scriptsize (0.50, 1.00) &  & \cellcolor{heatmapgreen!66} \textbf{0.800} \scriptsize (0.67, 1.00) & \cellcolor{heatmapgreen!44} 0.667 \scriptsize (0.50, 1.00) & \cellcolor{heatmapgreen!66} \textbf{0.800} \scriptsize (0.67, 1.00) \\
LaMe & 0.333 \scriptsize (1.00, 0.20) & & \cellcolor{heatmapgreen!20} 0.471 \scriptsize (0.57, 0.40) & \cellcolor{heatmapgreen!25} 0.500 \scriptsize (0.50, 0.50) & \cellcolor{heatmapgreen!49} 0.667 \scriptsize (0.53, 0.90) & \cellcolor{heatmapgreen!25} 0.500 \scriptsize (0.67, 0.40) &  & \cellcolor{heatmapgreen!45} 0.636 \scriptsize (0.58, 0.70) & \cellcolor{heatmapgreen!70} \textbf{0.800} \scriptsize (0.67, 1.00) & \cellcolor{heatmapgreen!33} 0.556 \scriptsize (0.62, 0.50) \\
PaPe & 0.600 \scriptsize (0.60, 0.60) & & \cellcolor{heatmappurple!4} 0.571 \scriptsize (1.00, 0.40) & \cellcolor{heatmappurple!16} 0.500 \scriptsize (0.43, 0.60) & \cellcolor{heatmapgreen!3} 0.615 \scriptsize (0.50, 0.80) & \cellcolor{heatmappurple!44} 0.333 \scriptsize (1.00, 0.20) &  & \cellcolor{heatmappurple!4} 0.571 \scriptsize (1.00, 0.40) & \cellcolor{heatmapgreen!50} \textbf{0.800} \scriptsize (0.80, 0.80) & \cellcolor{heatmappurple!4} 0.571 \scriptsize (1.00, 0.40) \\
PrDE & 0.333 \scriptsize (0.25, 0.50) & & \cellcolor{heatmappurple!33} 0.222 \scriptsize (0.33, 0.17) & \cellcolor{heatmapgreen!12} 0.417 \scriptsize (0.28, 0.83) & \cellcolor{heatmappurple!17} 0.276 \scriptsize (0.17, 0.67) & \cellcolor{heatmappurple!39} 0.200 \scriptsize (0.25, 0.17) &  & \cellcolor{heatmapgreen!33} \textbf{0.556} \scriptsize (0.42, 0.83) & \cellcolor{heatmapgreen!25} 0.500 \scriptsize (0.36, 0.83) & \cellcolor{heatmappurple!0} 0.333 \scriptsize (0.33, 0.33) \\
SeVD & 0.222 \scriptsize (0.25, 0.20) & & \cellcolor{heatmappurple!100} 0.000 \scriptsize (0.00, 0.00) & \cellcolor{heatmapgreen!22} 0.400 \scriptsize (0.40, 0.40) & \cellcolor{heatmapgreen!22} 0.400 \scriptsize (0.30, 0.60) & \cellcolor{heatmapgreen!14} 0.333 \scriptsize (1.00, 0.20) &  & \cellcolor{heatmapgreen!14} 0.333 \scriptsize (1.00, 0.20) & \cellcolor{heatmapgreen!44} \textbf{0.571} \scriptsize (1.00, 0.40) & \cellcolor{heatmapgreen!8} 0.286 \scriptsize (0.50, 0.20) \\
TrVD & 0.381 \scriptsize (0.27, 0.67) & & \cellcolor{heatmappurple!100} 0.000 \scriptsize (0.00, 0.00) & \cellcolor{heatmapgreen!7} 0.429 \scriptsize (0.38, 0.50) & \cellcolor{heatmappurple!17} 0.316 \scriptsize (0.23, 0.50) & \cellcolor{heatmapgreen!35} 0.600 \scriptsize (0.75, 0.50) &  & \cellcolor{heatmapgreen!46} \textbf{0.667} \scriptsize (0.67, 0.67) & \cellcolor{heatmapgreen!24} 0.533 \scriptsize (0.44, 0.67) & \cellcolor{heatmapgreen!35} 0.600 \scriptsize (0.75, 0.50) \\
\midrule
mean & 0.335 \scriptsize (0.39, 0.42) & & \cellcolor{heatmappurple!28} 0.241 \scriptsize (0.35, 0.19) & \cellcolor{heatmapgreen!9} 0.398 \scriptsize (0.33, 0.56) & \cellcolor{heatmapgreen!7} 0.388 \scriptsize (0.28, 0.76) & \cellcolor{heatmapgreen!6} 0.376 \scriptsize (0.58, 0.37) &  & \cellcolor{heatmapgreen!27} 0.518 \scriptsize (0.58, 0.58) & \cellcolor{heatmapgreen!36} \textbf{0.580} \scriptsize (0.53, 0.80) & \cellcolor{heatmapgreen!20} 0.469 \scriptsize (0.55, 0.45) \\

%% file: tables/decisiveness.tex
AdCO & & \cellcolor{heatmapgreen!100} 1.000 & \cellcolor{heatmapgreen!16} 0.160 & \cellcolor{heatmapgreen!16} 0.160 & \cellcolor{heatmapgreen!2} 0.023 &  & \cellcolor{heatmapgreen!99} 0.996 & \cellcolor{heatmapgreen!99} 0.992 & \cellcolor{heatmapgreen!1} 0.012 \\
AdVD & & \cellcolor{heatmapgreen!99} 0.993 & \cellcolor{heatmapgreen!12} 0.128 & \cellcolor{heatmapgreen!16} 0.164 & \cellcolor{heatmapgreen!0} 0.007 &  & \cellcolor{heatmapgreen!94} 0.947 & \cellcolor{heatmapgreen!100} 1.000 & \cellcolor{heatmapgreen!2} 0.023 \\
AdVO & & \cellcolor{heatmapgreen!100} 1.000 & \cellcolor{heatmapgreen!6} 0.066 & \cellcolor{heatmapgreen!8} 0.085 & \cellcolor{heatmapgreen!1} 0.011 &  & \cellcolor{heatmapgreen!99} 0.993 & \cellcolor{heatmapgreen!99} 0.996 & \cellcolor{heatmapgreen!2} 0.022 \\
DiCO & & \cellcolor{heatmapgreen!98} 0.988 & \cellcolor{heatmapgreen!31} 0.312 & \cellcolor{heatmapgreen!36} 0.362 & \cellcolor{heatmapgreen!5} 0.050 &  & \cellcolor{heatmapgreen!100} 1.000 & \cellcolor{heatmapgreen!100} 1.000 & \cellcolor{heatmapgreen!3} 0.037 \\
LaMe & & \cellcolor{heatmapgreen!99} 0.995 & \cellcolor{heatmapgreen!14} 0.145 & \cellcolor{heatmapgreen!8} 0.085 & \cellcolor{heatmapgreen!4} 0.040 &  & \cellcolor{heatmapgreen!90} 0.905 & \cellcolor{heatmapgreen!100} 1.000 & \cellcolor{heatmapgreen!4} 0.045 \\
PaPe & & \cellcolor{heatmapgreen!100} 1.000 & \cellcolor{heatmapgreen!6} 0.065 & \cellcolor{heatmapgreen!31} 0.315 & \cellcolor{heatmapgreen!0} 0.009 &  & \cellcolor{heatmapgreen!98} 0.981 & \cellcolor{heatmapgreen!100} 1.000 & \cellcolor{heatmapgreen!4} 0.046 \\
PrDE & & \cellcolor{heatmapgreen!99} 0.997 & \cellcolor{heatmapgreen!11} 0.115 & \cellcolor{heatmapgreen!10} 0.102 & \cellcolor{heatmapgreen!1} 0.010 &  & \cellcolor{heatmapgreen!89} 0.895 & \cellcolor{heatmapgreen!98} 0.990 & \cellcolor{heatmapgreen!2} 0.028 \\
SeVD & & \cellcolor{heatmapgreen!98} 0.989 & \cellcolor{heatmapgreen!5} 0.053 & \cellcolor{heatmapgreen!42} 0.421 & \cellcolor{heatmapgreen!1} 0.011 &  & \cellcolor{heatmapgreen!98} 0.989 & \cellcolor{heatmapgreen!94} 0.947 & \cellcolor{heatmapgreen!2} 0.021 \\
TrVD & & \cellcolor{heatmapgreen!100} 1.000 & \cellcolor{heatmapgreen!6} 0.060 & \cellcolor{heatmapgreen!26} 0.263 & \cellcolor{heatmapgreen!3} 0.030 &  & \cellcolor{heatmapgreen!100} 1.000 & \cellcolor{heatmapgreen!100} 1.000 & \cellcolor{heatmapgreen!3} 0.030 \\
\midrule
mean & & \cellcolor{heatmapgreen!99} 0.996 & \cellcolor{heatmapgreen!12} 0.123 & \cellcolor{heatmapgreen!21} 0.218 & \cellcolor{heatmapgreen!2} 0.021 &  & \cellcolor{heatmapgreen!96} 0.967 & \cellcolor{heatmapgreen!99} 0.992 & \cellcolor{heatmapgreen!2} 0.029 \\

%% file: tables/consistency_short.tex
\scriptsize 1-to-1 & & \cellcolor{heatmapgreen!12} 0.141 \scriptsize (0.23, 0.12) &  & \emph{N/A} \\
\scriptsize 1-to-N & & \cellcolor{heatmapgreen!53} 0.074 \scriptsize (0.09, 0.10) &  & \cellcolor{heatmapgreen!80} 0.031 \scriptsize (0.07, 0.03) \\
\scriptsize N-to-1 & & \cellcolor{heatmapgreen!61} 0.062 \scriptsize (0.05, 0.12) &  & \cellcolor{heatmapgreen!77} 0.037 \scriptsize (0.05, 0.06) \\
\scriptsize N-to-M & & \cellcolor{heatmapgreen!0} 0.160 \scriptsize (0.23, 0.18) &  & \cellcolor{heatmapgreen!41} 0.094 \scriptsize (0.10, 0.10) \\

%% file: sections/complementarity.tex
It is rare to find matching methods that combine high recall with high
precision. Since in practical data integration scenarios one needs to manually
verify the match candidates that are proposed by an automated matching
algorithm, its preferable from a practical viewpoint to use a matching algorithm
that has very high recall (to ensure that no candidates are missed) while
featuring a decent precision (to ensure that the verification effort remains
manageable).  From Table~\ref{tab:f1_scores}, we observe that using LLMs often
enhances recall compared to the baseline, with this improvement being more
pronounced for \texttt{GPT-4} than for \texttt{GPT-3.5}. Given this observation,
we investigate in this section how \emph{complementary} the different tasks
scopes are with the baseline and each other. For, if the sets of matching
candidates returned by distinct methods $A$ and $B$ are largely complementary (in
the sense that there is little overlap between the returned sets), we could
further increase recall by combining 
the methods $A$ and $B$ into a
method $A \& B$: the combined method simply returns the union of the matches of
$A$ and $B$. We must take care, however, as 
while the recall of $A \& B$ may increase compared to $A$ and $B$ alone, its precision will almost certainly decrease. As such, we are also interested in quantifying whether the  verification effort for $A \&B$ remains reasonable.



Our results in this section are computed using the following
methodology. We refer to an element of $\{ \baseline$, \texttt{1-1}, \texttt{1-N}, \texttt{N-1}, \texttt{N-M} $\}$ as a \emph{method}.
Remember from Section~\ref{sec:experimental_setup} that per method
we have repeated each experiment five times. Consequently per pair
$(S_1, S_2)$ of distinct methods we have $25$ experiment pairs $(E_1,
    E_2)$. We take the union of the matches resulting from $E_1$ and $E_2$ and
analyze this combined match w.r.t. the number of true positives, the recall, precision, etc. Per pair of methods $(S_1,S_2)$ we may compute a dataset-specific average of these methods by summing the metric result over all 25 experiment pairs, and taking the average.
Importantly, we only combine methods using the same LLM model (i.e. both use \texttt{GPT-3.5} or both use \texttt{GPT-4}). Concretely, in Section~\ref{sec:counts-semantic-matches} we analyze complementarity of matches by investigating how many additional true positive semantic matches may be recovered when combining methods. In Section~\ref{sec:verification_effort} we offset study by the verification effort required when combining methods. Finally, in
Section~\ref{sec:F1-scores-combined}, we analyze the F1-scores for every method
combination. 

\subsubsection{Counts of true semantic matches}%
\label{sec:counts-semantic-matches}

\begin{table}[tbp]
    \centering
    \caption{True semantic matches found when combining methods. A darker shade of green indicates a higher count. The diagonal shows the average count of true semantic	matches found by only that method.
        Recall that the ground truth consists of 49 matches.}
    \label{tab:combined_scopes}
    \footnotesize
    \setlength{\tabcolsep}{0.25em}
    \begin{tabular}{l c r c c c c r c c c}
                          &                         &  & \multicolumn{4}{c}{GPT-3.5} &                    & \multicolumn{3}{c}{GPT-4}                                                                                        \\
        \scriptsize scope & \scriptsize $\baseline$ &  & \scriptsize 1-to-1          & \scriptsize 1-to-N & \scriptsize N-to-1        & \scriptsize N-to-M &  & \scriptsize 1-to-N & \scriptsize N-to-1 & \scriptsize N-to-M \\
        \midrule
        \input{tables/combinations_gpt35.tex}
    \end{tabular}
\end{table}

Table~\ref{tab:combined_scopes} shows the number of true semantic matches (i.e., true positives) found by combining methods. Concretely, for method combination $(i,j)$ and each dataset we first compute the average number of true positives returned. We then sum these averages over all datasets, and report this sum in cell $(i,j)$. The diagonals  show the average
count of true semantic matches found by the corresponding method alone, thus not combined with another method. This number serves as a reference
for the method combinations: numbers for method combinations that are higher indicate an increase compared to using the method alone.

We first discuss the findings for \texttt{GPT-3.5} and then those for
\texttt{GPT-4}. Remember that there is a total of 49 true semantic matches in the ground truth (cf.\ Table~\ref{tab:datasets}).




We observe that combining \texttt{1-to-N} and \texttt{N-to-1} yields the
highest count of true semantic matches on average (40 out of 49). \texttt{N-to-1} by itself uncovers most true semantic matches on
average ($35.4$ for \texttt{GPT-3.5}), which makes it the best task scope to
combine with. We see that any combination with \texttt{N-to-1} yields more
matches than any combination without \texttt{N-to-1}. As such, \texttt{N-to-1} is complementary with all other methods. This observation even
holds when using the larger \texttt{GPT-4} model, even though the number of
semantic matches found by \texttt{N-to-1} is higher on average than with
\texttt{GPT-3.5}. The best combination of task scopes without \texttt{N-to-1}
is $\baseline$ combined with \texttt{1-to-N}, yielding an average count of $32.4$. We note that this value is worse than using \texttt{N-to-1} on its own.
Overall, \texttt{1-to-N} is the second-best combination partner, as every
average counts gets lower if we swap out \texttt{1-to-N} for any other task
scope except \texttt{N-to-1}. The use of \texttt{GPT-4} does not notably
improve the average counts of true semantic matches for combinations with
\texttt{1-to-N} or \texttt{N-to-1}. 
With \texttt{N-to-M}, however, we do see an
increase for all combinations. 

\paragraph{Conclusion.} Combining \texttt{1-to-N} and \texttt{N-to-1} yields
the highest count of true semantic matches on average. The use of
\texttt{GPT-4} does not improve this count by much.

\subsubsection{Verification effort required}%
\label{sec:verification_effort}

\begin{table}[tbp]
    \centering
    \caption{Verification effort: the count of matches that have to be inspected when combining two task scopes. The diagonals show the average matches found by a single task scope. A darker green indicates less effort. Recall that our total search space consists of $1839$ attribute pairs and there are 49 matches in the ground truth.}
    \label{tab:combined_verification_effort}
    \footnotesize
    \setlength{\tabcolsep}{0.25em}
    \begin{tabular}{l c r c c c c r c c c}
                          &                         &  & \multicolumn{4}{c}{GPT-3.5} &                    & \multicolumn{3}{c}{GPT-4}                                                                                        \\
        \scriptsize scope & \scriptsize $\baseline$ &  & \scriptsize 1-to-1          & \scriptsize 1-to-N & \scriptsize N-to-1        & \scriptsize N-to-M &  & \scriptsize 1-to-N & \scriptsize N-to-1 & \scriptsize N-to-M \\
        \midrule
        \input{tables/combined_verification_effort.tex}
    \end{tabular}
\end{table}

To assess the human effort required to verify candidate matches we report in
Table~\ref{tab:combined_verification_effort} the size of the match candidates returned per combined method. Concretely, for method combination $(i,j)$ and each dataset we first compute the average cardinality of the set of returned match candidates. We then sum these averages over all datasets, and report this sum in cell $(i,j)$. The diagonals show the cardinality of the corresponding method alone, thus not combined with another method.

A larger cell value means that more candidates
need to be inspected. Remember from Table~\ref{tab:datasets} that the search
space of possible matchings our benchmark consists of a total of $1839$ attribute combinations and there are 49 true semantics matches in the ground truth. We
include the verification effort for \texttt{1-to-1} for completeness, but
do not discuss it in detail because of the low result quality.

Overall, we deem most LLM-counts acceptable for manual verification as they are
much smaller ($<10\%$) than the entire search space of all attribute pairs. Looking specifically at methods used in isolation (shown on the diagonal) we
observe that with \texttt{GPT-3.5}, \texttt{N-to-1} retrieves the most matches
($136.2$), followed by \texttt{1-to-N} ($104.6$) and \texttt{N-to-M} ($30.0$).
Using \texttt{GPT-4}, the counts of \texttt{1-to-N} and \texttt{N-to-1} are further reduced while the count of \texttt{N-to-M} increases.

When we combine two task scopes, we observe that \texttt{1-to-N} combined with
\texttt{N-to-1} retrieves the highest number of match candidates, $183.6$ to be
precise. We deem this acceptable for practical applications as it represents
roughly only $10\%$ of our entire search space. We note that while $183.6$ candidates to inspect may still seem a lot, these numbers are aggregated over all datasets in our benchmark. When drilling down to the dataset level there are on average fewer than 20 candidates to verify using \texttt{GPT-3.5} (often much less), which reduces to fewer than 10 candidates to verify using \texttt{GPT-4}. Compared to the number of possible pairs per dataset shown in Table~\ref{tab:datasets} this remains very modest.



\paragraph{Conclusion.}
Our experiments show that the number of retrieved matches is very reasonable
($<10\%$ of the search space) and can be reduced further with the use of
\texttt{GPT-4}, rendering the verification effort for all task scope
combinations acceptable.


\subsubsection{F1-scores}
\label{sec:F1-scores-combined}

Table~\ref{tab:combined_f1_scores} presents F1-scores, precision and
recall for all method combinations, averaged over all datasets. The
diagonals represent the average scores for a single task scope and can be
roughly compared to the last row of Table~\ref{tab:f1_scores}, where we report
the average of the median scores. The table is meant to be read row-wise: the cell in row $i$ and column $j$ shows how combining method $i$ with method $j$ behaves compared to using method $i$ alone.

First, let us discuss combining $\baseline$ with the different task
scopes (first row of Table~\ref{tab:combined_f1_scores}). We observe that any
combination of $\baseline$ with any task scope increases the F1-score compared to
using $\baseline$ alone own. Using \texttt{GPT-3.5}, \texttt{1-to-1} achieves the highest improvement,
yielding an F1-score of $0.384$ on average; \texttt{N-to-1} has the lowest
improvement with $0.344$. We see that the low F1-score of \texttt{N-to-1} can
be attributed to a low precision, as recall is highest across all combinations
including $\baseline$. Using \texttt{GPT-4} increases the F1-scores further,
yielding a maximum F1-score of $0.456$ in combination with \texttt{N-to-1}.
These numbers, hence, show how LLMs strictly improve over
string-similarity-based matching.

Overall, we see that combining two task scopes improves the F1-score. Using
\texttt{GPT-3.5}, we see three exceptions: ($\baseline$, \texttt{N-to-1}),
where using \texttt{N-to-1} on its own yields a higher F1-score;
(\texttt{1-to-N}, \texttt{N-to-1}), where the F1-score is lower than using any
task scope on its own; and (\texttt{1-to-N}, \texttt{N-to-M}), where using
\texttt{1-to-N} on its own yields a higher F1-score. Further, combining
\texttt{1-to-N} or \texttt{N-to-1} with $\baseline$ also reduces the F1-score
while it improves for \texttt{1-to-1} and \texttt{N-to-M}. We also observe that
the combination (\texttt{1-to-N}, \texttt{N-to-1}) achieves the highest recall
of all combinations, even including the ones using \texttt{GPT-4}. Using the
larger model, we see a similar trend as with \texttt{GPT-3.5} in that a task
scope combination typically improves the average F1-score while combining with
$\baseline$ worsens it. Further, combining \texttt{N-to-1} with any other task
scope reduces its F1-score as well, making it the highest performing task scope
based on the average F1-score. Looking at the \texttt{GPT-4} experiments
in isolation, we again observe that the combination (\texttt{1-to-N},
\texttt{N-to-1}) achieves the best recall on average.

We note that while the F1-scores are generally not very high, we see that
the task scopes achieve very different scores for precision and recall. Any
combination with \texttt{N-to-1} generally improves recall at the cost of
precision. For \texttt{1-to-N} the gap between precision and recall is
similar but less pronounced. In contrast, both \texttt{1-to-1} and
\texttt{N-to-M} do not contribute much to recall while keeping precision level.

\paragraph{Conclusion.} Combining task scopes generally improves the F1-score
on most combinations, with \texttt{N-to-1} using \texttt{GPT-4} achieving the
highest F1-score. 

\begin{table*}[tbp]
    \centering
    \caption{F1-scores, precision and recall  for combined methods, averaged over all datasets. The diagonals represent the average scores for a single task scope and provide the reference point for row-wise comparisons. A green colouring indicates a higher F1-score compared using the method mentioned in the row on its own, purple indicates a lower F1-score. Note that the precision and recall scores are also averages and thus do not directly correspond to the F1-scores shown.}
    \label{tab:combined_f1_scores}
    \footnotesize
    \setlength{\tabcolsep}{0.15em}
    \begin{tabular}{l c r c c c c r c c c}
                          &                         &  & \multicolumn{4}{c}{GPT-3.5} &                    & \multicolumn{3}{c}{GPT-4}                                                                                        \\
        \scriptsize scope & \scriptsize $\baseline$ &  & \scriptsize 1-to-1          & \scriptsize 1-to-N & \scriptsize N-to-1        & \scriptsize N-to-M &  & \scriptsize 1-to-N & \scriptsize N-to-1 & \scriptsize N-to-M \\
        \midrule
        \input{tables/combined_f1_scores.tex}
    \end{tabular}
\end{table*}

%% file: tables/combinations_gpt35.tex
\scriptsize $\baseline$ & \cellcolor{heatmapgreen!0}19.0 & & \cellcolor{heatmapgreen!60}24.2 & \cellcolor{heatmapgreen!81}32.4 & \cellcolor{heatmapgreen!94}37.8 & \cellcolor{heatmapgreen!60}24.2 & & \cellcolor{heatmapgreen!79}31.6 & \cellcolor{heatmapgreen!96}38.4 & \cellcolor{heatmapgreen!71}28.6 \\
\scriptsize 1-to-1 & & & \cellcolor{heatmapgreen!0}9.8 & \cellcolor{heatmapgreen!72}29.1 & \cellcolor{heatmapgreen!91}36.6 & \cellcolor{heatmapgreen!51}20.4 & & \emph{N/A} & \emph{N/A} & \emph{N/A} \\
\scriptsize 1-to-N & & & & \cellcolor{heatmapgreen!0}27.0 & \cellcolor{heatmapgreen!100}40.0 & \cellcolor{heatmapgreen!73}29.2 & & \cellcolor{heatmapgreen!0}28.2 & \cellcolor{heatmapgreen!98}39.4 & \cellcolor{heatmapgreen!73}29.6 \\
\scriptsize N-to-1 & & & & & \cellcolor{heatmapgreen!0}35.4 & \cellcolor{heatmapgreen!93}37.2 & & & \cellcolor{heatmapgreen!0}38.4 & \cellcolor{heatmapgreen!96}38.5 \\
\scriptsize N-to-M & & & & & & \cellcolor{heatmapgreen!0}14.4 & & & & \cellcolor{heatmapgreen!0}21.8 \\

%% file: tables/combined_verification_effort.tex
\scriptsize $\baseline$ & \cellcolor{heatmapgreen!65}77.0 & & \cellcolor{heatmapgreen!58}93.2 & \cellcolor{heatmapgreen!29}159.0 & \cellcolor{heatmapgreen!17}185.6 & \cellcolor{heatmapgreen!58}94.0 & & \cellcolor{heatmapgreen!48}116.2 & \cellcolor{heatmapgreen!40}134.2 & \cellcolor{heatmapgreen!54}103.4 \\
\scriptsize 1-to-1 & & & \cellcolor{heatmapgreen!90}21.8 & \cellcolor{heatmapgreen!50}110.6 & \cellcolor{heatmapgreen!37}141.0 & \cellcolor{heatmapgreen!79}45.8 & & \emph{N/A} & \emph{N/A} & \emph{N/A} \\
\scriptsize 1-to-N & & & & \cellcolor{heatmapgreen!53}104.6 & \cellcolor{heatmapgreen!18}183.6 & \cellcolor{heatmapgreen!50}112.2 & & \cellcolor{heatmapgreen!71}63.8 & \cellcolor{heatmapgreen!56}96.8 & \cellcolor{heatmapgreen!68}69.8 \\
\scriptsize N-to-1 & & & & & \cellcolor{heatmapgreen!39}136.2 & \cellcolor{heatmapgreen!36}142.4 & & & \cellcolor{heatmapgreen!62}84.0 & \cellcolor{heatmapgreen!61}86.4 \\
\scriptsize N-to-M & & & & & & \cellcolor{heatmapgreen!86}30.0 & & & & \cellcolor{heatmapgreen!81}42.6 \\

%% file: tables/combined_f1_scores.tex
\scriptsize $\baseline$ & \cellcolor{heatmappurple!0}0.335 \scriptsize (0.39, 0.42) & & \cellcolor{heatmapgreen!24}0.384 \scriptsize (0.34, 0.53) & \cellcolor{heatmapgreen!12}0.359 \scriptsize (0.27, 0.69) & \cellcolor{heatmapgreen!4}0.344 \scriptsize (0.24, 0.78) & \cellcolor{heatmapgreen!20}0.375 \scriptsize (0.35, 0.53) & & \cellcolor{heatmapgreen!35}0.406 \scriptsize (0.32, 0.63) & \cellcolor{heatmapgreen!60}0.456 \scriptsize (0.35, 0.78) & \cellcolor{heatmapgreen!36}0.409 \scriptsize (0.34, 0.59) \\
\scriptsize 1-to-1 & & & \cellcolor{heatmappurple!0}0.234 \scriptsize (0.35, 0.19) & \cellcolor{heatmapgreen!91}0.417 \scriptsize (0.35, 0.62) & \cellcolor{heatmapgreen!72}0.378 \scriptsize (0.27, 0.76) & \cellcolor{heatmapgreen!95}0.425 \scriptsize (0.49, 0.44) & & \emph{N/A} & \emph{N/A} & \emph{N/A} \\
\scriptsize 1-to-N & & & & \cellcolor{heatmappurple!0}0.406 \scriptsize (0.35, 0.59) & \cellcolor{heatmappurple!27}0.351 \scriptsize (0.23, 0.83) & \cellcolor{heatmappurple!3}0.400 \scriptsize (0.33, 0.64) & & \cellcolor{heatmappurple!0}0.505 \scriptsize (0.54, 0.58) & \cellcolor{heatmapgreen!21}0.547 \scriptsize (0.46, 0.80) & \cellcolor{heatmapgreen!4}0.513 \scriptsize (0.50, 0.61) \\
\scriptsize N-to-1 & & & & & \cellcolor{heatmappurple!0}0.373 \scriptsize (0.27, 0.73) & \cellcolor{heatmapgreen!7}0.388 \scriptsize (0.27, 0.78) & & & \cellcolor{heatmappurple!0}0.572 \scriptsize (0.50, 0.78) & \cellcolor{heatmappurple!4}0.562 \scriptsize (0.47, 0.78) \\
\scriptsize N-to-M & & & & & & \cellcolor{heatmappurple!0}0.355 \scriptsize (0.49, 0.35) & & & & \cellcolor{heatmappurple!0}0.486 \scriptsize (0.55, 0.48) \\

%% file: sections/conclusion.tex
In this study, we took an initial step towards utilizing LLMs for schema matching. We found that matching quality diminishes when there is insufficient context information (i.e., task scope \texttt{1-to-1}) and when there is an excess of context information (i.e., task scope \texttt{N-to-M}). The latter is likely hindered by the more complex output format and the larger number of pairs requiring decisions. The \texttt{1-to-N} and \texttt{N-to-1} task scopes effectively  provide sufficient context to make accurate matches without overwhelming the decision-making process.
This balance results in a better overall performance of which the recall can be even further enhanced by adopting a combined approach using both task scopes in tandem. This combined method successfully identifies a significant number of true semantic matches with an acceptable verification effort. As such, we recommend using the combined (\texttt{1-to-N}, \texttt{N-to-1}) method in practice.
We also found that using GPT-4 over GPT-3.5
improves matching quality and consistency over all task scopes tested on both models, and (except for \texttt{N-to-M}) increases decisiveness and reduces the verification effort. 
The results in this paper demonstrate that LLMs have the potential to bootstrap the schema matching process and assist data engineers in speeding up this task solely based on schema element names and descriptions, without the need for data instances and improving over attribute-name-based matching alone.

We outline some directions for future work that seem promising.

A benefit of LLMs over the string similarity baseline is that they can
be instructed to provide an explanation as to why they identify a certain attribute pair as a match or a non-match. We believe that such explanations can be a valuable instrument for a data engineer tasked to construct a schema mapping, to identify and rectify misclassifications.
Through initial experiments, we have observed that the LLM sometimes jumps to conclusions as it overemphasizes similarity of attribute names while disregarding the intent of the attributes as described in the provided documentation. For instance, we noticed that the LLM is eager to match two attributes solely based on the fact that they both refer to the time dimension of an event even when those events are clearly different.
We are currently working on a tool that facilitates refining schema matchings via natural language feedback in a pragmatic and user-friendly way.


Our benchmark consists of publicly available schemas.
In future experiments, we will apply our approach on proprietary schemas, aiming to illustrate the usefulness of using LLMs for schema matching in real-world scenarios.
